\documentclass[reprint,superscriptaddress,aps,longbibliography]{revtex4-2}
\usepackage[utf8]{inputenc}
\usepackage[T1]{fontenc}
\usepackage{amsmath}
\usepackage{amssymb}
\usepackage{graphicx}
\usepackage[capitalize, nameinlink]{cleveref}
\usepackage{subfigure}
\graphicspath{{figures/}}
\usepackage{adjustbox}
\usepackage{booktabs}
\usepackage{multirow}

\usepackage{color}
\usepackage{rotating}
\usepackage{ulem}

\begin{document}

\title{Generation of spin wave packets by reconfigurable magnonic heterojunctions}

\author{A. Roxburgh}
\affiliation{Center for Magnetism and Magnetic Nanostructures, University of Colorado Colorado Springs, Colorado Springs, CO, USA}
\author{P. Micaletti}
\affiliation{ Dipartimento di Fisica e Scienze della Terra, Universit\`a di Ferrara, Ferrara, Italy }
\author{F. Montoncello}
\affiliation{ Dipartimento di Fisica e Scienze della Terra, Universit\`a di Ferrara, Ferrara, Italy }
\author{E. Iacocca}
\email{eiacocca@uccs.edu}
\affiliation{Center for Magnetism and Magnetic Nanostructures, University of Colorado Colorado Springs, Colorado Springs, CO, USA}

\date{\today}

\begin{abstract}
Nano-magnonic crystals are magnetic waveguides whose magnetic parameters are modulated at the nanoscale. The super-lattice structure enables a band structure and magnonic band-gaps. Here, we numerically investigate the field tunability of such a magnonic band-gap. By subjecting different parts of the nano-magnonic crystal to a field, we realize a magnonic heterojunction. A numerical demonstration of an active modification of the band structure leads to magnonic amplitude modulation in the linear regime and magnonic frequency combs in the nonlinear regime. Our results offer further opportunities for nano-magnonic crystals and incite their experimental realization.
\end{abstract}
\maketitle

\section{Introduction}

The investigation and control of the quanta of angular momentum, or magnonics, has found many potential applications including storage, data processing, information transport, and progress towards magnon-based quantum computing~\cite{Chumak2022, Barman_2021}. An interesting branch of magnonics explores reconfigurable structures~\cite{Grundler2015, Kumar2014}, that is, devices whose properties can be actively tuned by external stimuli. For example, periodic patterning of magnetic materials establishes a magnonic crystal~\cite{Krawczyk2014} exhibiting magnonic band gaps~\cite{Wang2009,Tacchi2012,Graczyk2018}. Other forms of periodic patterning are possible such as modulating the Dzyaloshinskii-Moriya interaction~\cite{Gallardo2019} and the magnetization state~\cite{Montoncello2021,Micaletti2024}, resulting in both band gaps and flat bands.

The reconfigurable control of magnons~\cite{Grundler2015} has been so far tightly related to the magnonic crystal features. For example, the magnetic state can be reconfigured to alter the magnon propagation characteristics~\cite{Haldar2016,Gubbiotti2018,Mondal2024}. Other forms of magnon control include the manipulation of the energy landscape~\cite{Karenowska2012} and the use of magnetic domain walls as waveguides~\cite{Wagner2016}. Reconfigurability of magnetization states can be also attained by vertical coupling with a patterned layer, e.g., an artificial spin ice (ASI) lattice~\cite{Iacocca2020,Kempinger2021,Negrello2022,Szulc2022}. For example, in ASI-film hybrid materials here, reconfiguring the ASI microstates~\cite{Arroo2019,Vanstone2022} makes it possible to control the magnetization distribution and hence the magnon propagation along the film layer.

A new frontier in the fabrication of magnonic crystals is experimentally achievable with a technique called thermal nano-lithography (TNL)~\cite{Albisetti2016, Levati2023,Barker2024}. This method utilizes an atomic force microscope with a tip that can be heated to 1,100$^{\circ}$C to perform local annealing of a material with nanoscale precision. Several magnetic parameters can be modified with this technique based on the material structure of a multilayer. It was recently shown that band gaps arise by modulating the exchange constant, $A$, with a period in the order of $50$~nm~\cite{Roxburgh2024}, which can be considered to be a nano-magnonic crystal. The band-gaps arise due to the fact that a super-lattice is established and wave coupling leads to the breaking of degeneracy, as is well known from the Bloch theorem in crystals. An important difference between this technique and traditional magnonic crystals to achieve a super-lattice is that the magnons are primarily exchange-mediated in our case. This implies that the super-lattice is both of nanoscale dimensions and easily reconfigurable by, e.g., external magnetic fields, providing a significant improvement in active magnon manipulation.

Here, we numerically investigate the realization of magnetic heterojunctions controllable by external magnetic fields. The basis for this structure is the nano-magnonic crystal in which a band-gap on the order of $21$~GHz is already present. A local magnetic field allows us to shift the band-gap, thus realizing a heterojunction. Temporal modulation of the external field in the MHz range gives rise to magnon wave packets. Because of the nonlinearity of this process, the resulting wave packet spectrum exhibits a frequency comb, a magnonic realization~\cite{Wang2021,Hula2022} more closely related to optical frequency combs~\cite{Fortier2019}. Wave-packets have been experimentally achieved by modulating the current in an antenna and optically detected~\cite{Heinz2020}.

The remainder of the paper is organized as follows. In section II, we discuss the tunability of band gaps due to a magnetic field as our numerically controllable parameter. This forms the basis for section III, where we split our sample into regions of differing fields to establish a heterojunction. We investigate the effect of this configuration as a transfer function for a magnon. In section IV, we demonstrate the appearance of wave packets as the band gap is modulated by varying field strengths, We also discuss the presence of frequency combs. Finally, we provide concluding remarks in section V.

\section{Geometry and magnonic band gaps}

We assume an elongated thin permalloy nano-magnonic crystal with a periodic modulation of the exchange constant. The modulation has a wavenumber $q$, along the same direction as the magnon wavenumber $k$. We then assume an external field $\mathbf{H}_0=H_o\hat{\mathbf{y}}$ applied along the plane of the material and perpendicular to the wavenumber $k$ so that the uniform magnetization is $m_y\approx1$. In this surface wave configuration, band-gaps arise~\cite{Roxburgh2024}. Here, we use material parameters for permalloy: $M_s=790$~kA/m, $A=10$~pJ/m, and $\alpha=0.01$. The periodic modulation has $q\approx0.16$~rad/nm, corresponding to a period of $40$~nm. The exchange constant is modulated by this period according to $A_{mod}=A[1+\beta\sin{(qx)}]$, with $\beta=0.4$. This value of $\beta$ was shown in Ref.~\cite{Roxburgh2024} to give rise to sizable band gaps, on the order of a few GHz. The external magnetic field is considered to be a tunable parameter that simply shifts the band structure.

\begin{figure}[t]
 \includegraphics[width=3.3in, trim={0 0.3in 0 0.5in},clip]{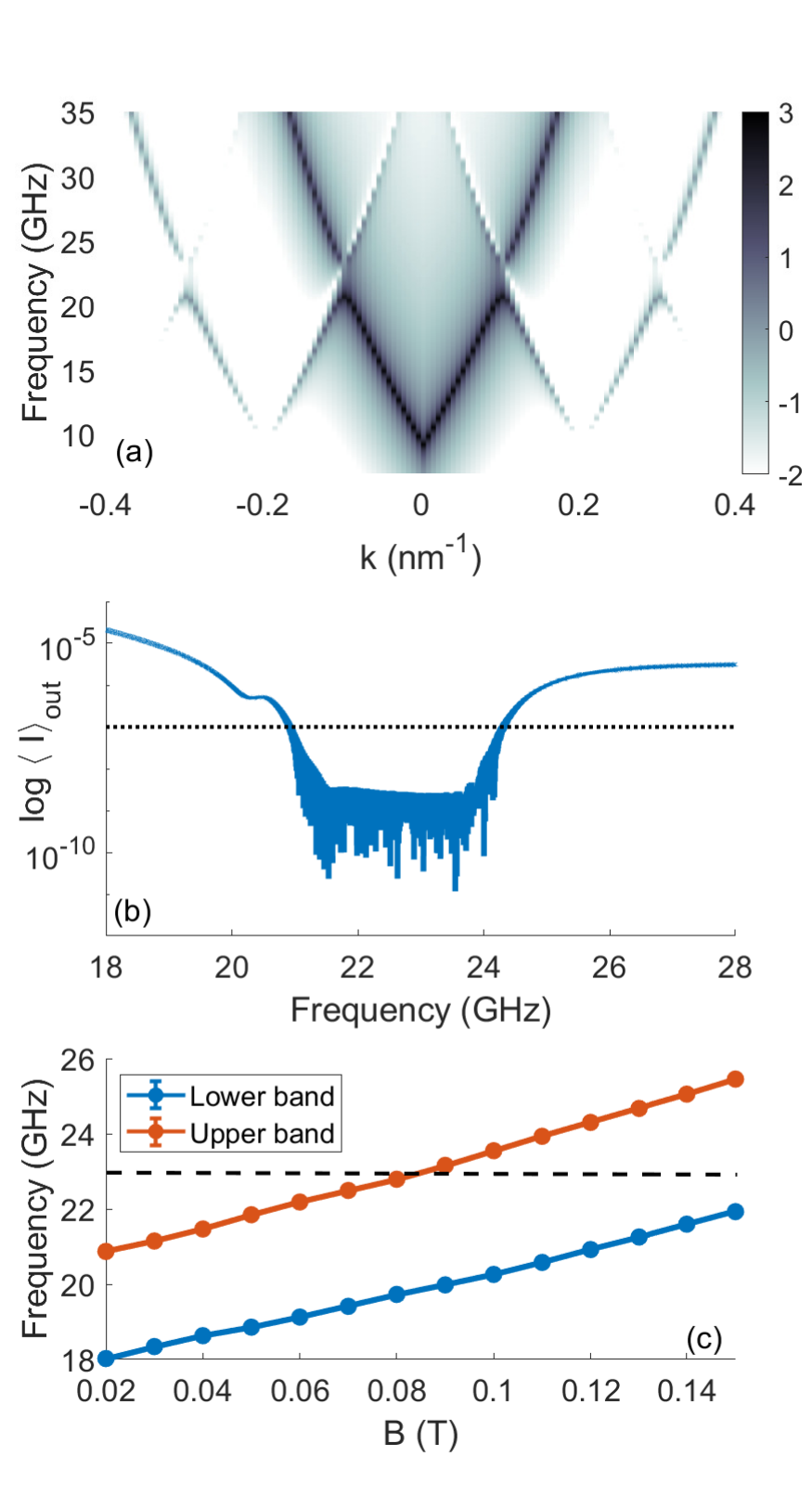}
 \caption{(a) Band structure of a nano-magnonic crystal with modulated exchange constant in the surface wave configuration and an applied field of $0.1$~T. A band-gap can be assumed at $~\approx 22$~GHz. (b) Quantification of the band-gap by use of a chirped magnon excitation. The band-gap is seen as a strong attenuation of the measured magnon a the other edge of the nano-magnonic crystal. The horizontal dashed black line indicates our tolerance level to measure the band-gap. (c) Band-gap tunability as a function of the external magnetic field displaying the expected linear behavior. The determined errorbars are smaller than the plotted symbols.}
 \label{fig:NoBarrier}
 \end{figure}

We create this configuration using MuMax3~\cite{Vansteenkiste2014}, a finite difference, GPU-accelerated micromagnetic package. The permalloy nano-magnonic crystal had dimensions $2000$~nm~$\times250$~nm~$\times5$~nm, with cell size of approximately $3.9$~nm~$\times7.8$~nm~$\times5.0$~nm based on a discretization of $512\times32\times1$ cells. Due to our choice of $q$, the primitive cell is 40-nm long, as mentioned earlier. Periodic boundary conditions were set in the y direction to ensure a homogeneous magnon profile. The band structure can be obtained by exciting the layer with a sinc function in space and time~\cite{Venkat2013}, where we limit the frequency axis to $35$~GHz. A resulting band structure for an external magnetic field magnitudes of 0.1~T is shown in Fig.~\ref{fig:NoBarrier}(a). Clearly, the modulation of the exchange constant creates a periodicity and band-gaps open. 

To quantify the band-gaps as a function of field, we excited magnons at one edge of the permalloy nano-magnonic crystal and collected the average magnetization in the $z$ component, $\langle m_z\rangle_{out}$ at the opposite edge. For numerical efficiency, magnons were excited by a local oscillating magnetic field with a linearly changing frequency, i.e., a chirp. Because the group velocity of magnons is  proportional to their wavevector, we swept the frequencies from their maximum value of $f_{high}=30$~GHz to its minimum at $f_{low}=15$~GHz. This allows waves to expand in time, thus avoiding magnon interactions. Because $\langle m_z\rangle_{out}$ has both positive and negative values, we numerically determine its envelope, $\langle I\rangle_{out}$, so that the output is definite positive. The choice of an envelope over a simple absolute value is based on the fact that the envelope filters the oscillations. This is equivalent to an amplitude modulation detection scheme. To plot $\langle I\rangle_{out}$ as a function of frequency, we map the simulation time to frequency by the equation
\begin{equation}
    \label{eq:freq}
    f = \frac{L^2}{4\gamma\mu_{0}M_{s}\lambda_{ex}^2t^2}+f_{high}-\frac{f_{high}-f_{low}}{t_{max}}t.
\end{equation}
where the first term is essentially a frequency shift related to the group velocity of waves to reach the farther end of the nano-magnonic crystal. To derive this term, we note that the time to reach the end of the nano-magnonic crystal is $t=L/v_g$, with $v_g=2\gamma\mu_0M_s\lambda_\mathrm{ex}^2k$ being the group velocity of exchange-dominated waves, for which the dispersion relation reads $\omega=\gamma\mu_0 M_s\lambda_{ex}^2 k^2$. The remaining terms in Eq.~\eqref{eq:freq} simply come from the chirp. Here, $L=2000$~nm the length of the nano-magnonic crystal, $\gamma=28$~GHz/T the reduced gyromagnetic ratio, $\mu_0$ the vacuum permeability, and $\lambda_{ex}=\sqrt{2A/\mu_0M_s^2}$ is the average exchange length, and $t$ is the simulation time. Note that Eq.~\eqref{eq:freq} diverges at $t=0$ which is reasonable since magnons have a finite group velocity. Therefore, this expression is only reasonable in the vicinity of the frequencies used in the chirp.

An example of $\langle I\rangle_{out}$ revealing the band gap frequency range is shown in Fig.~\ref{fig:NoBarrier}(b) for an external field $B=0.12$~T by the noticeable dip. In other words, this is a region where magnons are evanescent and only noise reaches the end of the permalloy nano-magnonic crystal. To quantify the band-gap, we set an arbitrary ``noise floor'' of $10^{-7}$, illustrated by a horizontal dashed black line. The lower and upper edges of the band-gap are obtained when the data crosses this line, allowing us to compute both a mean value and standard deviation. The results are shown in Fig.~\ref{fig:NoBarrier}(c), displaying a linear tunability the band-gap from $B=0.02$~T to $B=0.15$~T.

\section{Magnonic heterojunction}

Because of the field tunability of the band-gap, a magnonic heterojunction can be established by applying a different external field in a section of the permalloy nano-magnonic crystal. This results in a misalignment of the band-gaps, an analogy to semiconducting heterojunctions. An important difference is that magnonic bands are aligned by the minimum energy rather than the work function and that the band-gap is tunable as opposed of material dependent.

\begin{figure}[t]
 \includegraphics[width=3.3in]{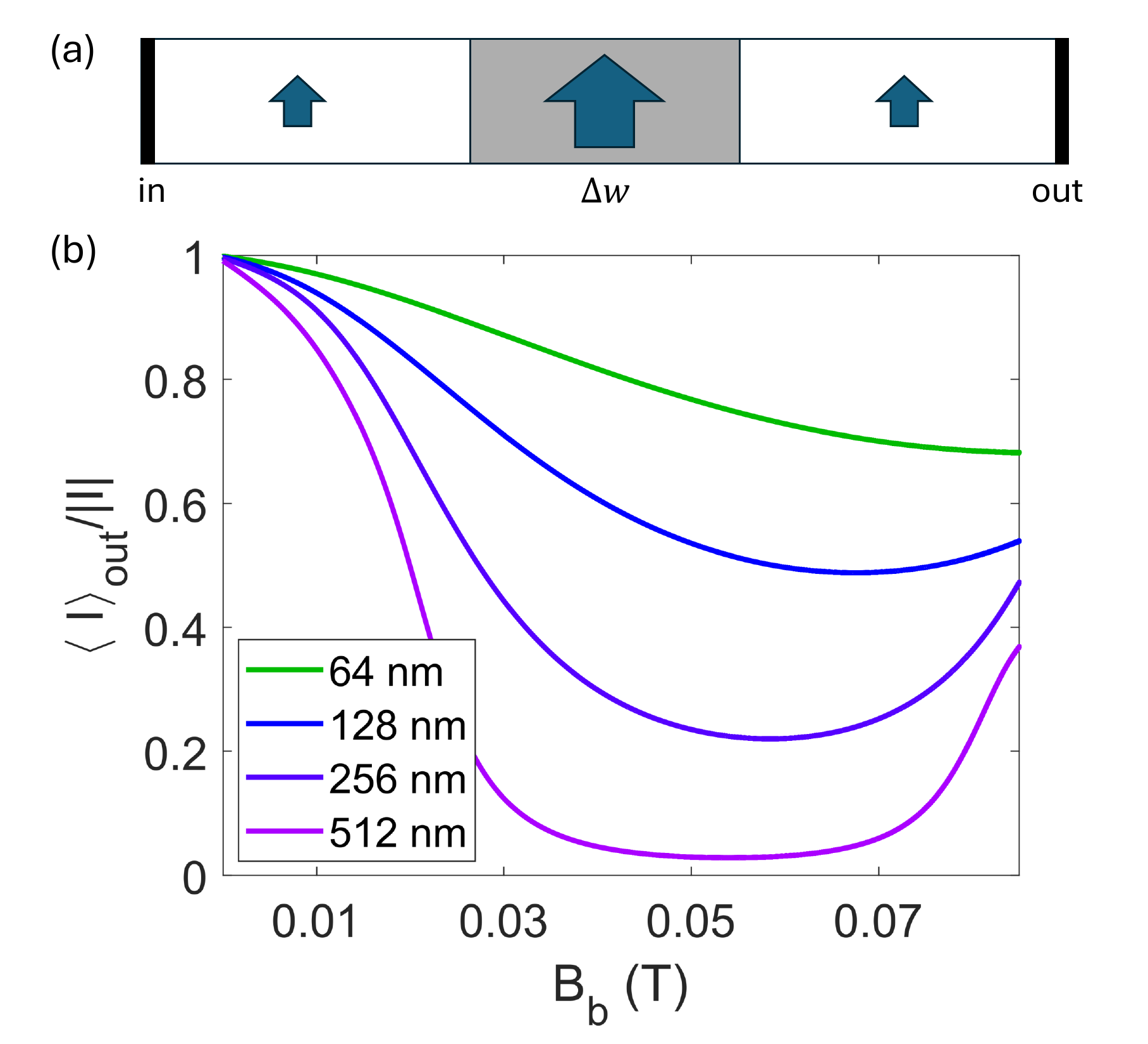}
 \caption{(a) Schematic of the magnonic heterojunction. The black areas represent the regions where the magnon is generated (``in'') and measured (``out''). The whole nano-magnonic crystal is subject to a magnetic field represented by blue arrows. In the gray region of width $\Delta w$, the magnetic field is varied by adding a field $B_b$. (b) Normalized output magnetization envelope as a function of $B_b$ and for several choices of $\Delta w$ at the fixed magnon frequency of $23$~GHz.}
 \label{fig:DynamicBarrier}
 \end{figure}

The schematic of our chosen configuration is shown in Fig.~\ref{fig:DynamicBarrier}(a), where the ``barrier'' in the middle region of width $\Delta w$ is subject to a different magnetic field, $B_b$. The left and right edges of our configuration are labeled ``in'' and ``out'' respectively, indicating the excitation and detection of magnons, as described previously.

Based on Fig.~\ref{fig:NoBarrier}(c), we chose a magnon frequency of $23$~GHz which propagates through the permalloy nano-magnonic crystal when $B<0.08$~T is applied. The field in the barrier can be raised to arrest the magnon propagation at this frequency. We characterize this behavior by linearly varying the field magnitude in the barrier from $0.05$~T to $0.15$~T in time. We assume here that the field transition is sharp to investigate the ideal behavior. Similar to Fig.~\ref{fig:NoBarrier}(b), the simulation time can be expressed as a field through a simple linear conversion. Because in this case we use a particular magnon frequency, we allow the magnon to stabilize for $20$~ns and there is temporal shift of $\approx 3$~ns due to the magnon group velocity. The stabilized magnon has an average output envelope of $|I|$ which is used below to normalize the field-dependent envelope.

The normalized output magnetization envelope $\langle I\rangle_{out}/|I|$ for several barrier widths are shown in Fig.~\ref{fig:DynamicBarrier}(b). We considered barriers of widths $64$~nm, $128$~nm, $256$~nm, and $512$~nm. Clearly, the band-gap dip becomes well-defined as the width increases. There are two reasons for this. First, magnons within the band-gap are evanescent and thus exponentially decay in space proportional to their frequency and Gilbert damping. Second, even though the spatial change in the field is sharp, the exchange interaction favors a smooth magnetization change. Thus, narrow barriers do not exhibit a fully formed band-gap. This can be considered to be analogous to the depletion region in p-n junctions. It must be noted that if the spatial change in the field is not sharp, as would be in an experimental case, the depletion region would also be smoothed out. However, this would not affect the main functionality of the barrier: in the region where the field is uniform, magnons would be evanescent.

Because of the aforementioned reasons, the widest barrier considered attenuates magnons by a factor $\approx33$. This attenuation is relative to the measured spin wave amplitude when the field is uniform throughout the nano-magnonic crystal. It must be pointed out that it might be difficult experimentally detect magnons in a $2000$~nm permalloy strip~\cite{Madami2011}. A possibility is to enhance magnon propagation by currents, e.g., by spin-Hall effect~\cite{Woo2017} or spin-transfer torque~\cite{Seo2009}. Another alternative that could prove useful is the use lower-damping materials, e.g., YIG~\cite{Serga2010,Collet2017,Frey2020} but TNL has yet to be demonstrated in insulating materials to the best of our knowledge.
\begin{figure*}[t]
\begin{center}
\includegraphics[width=6in]{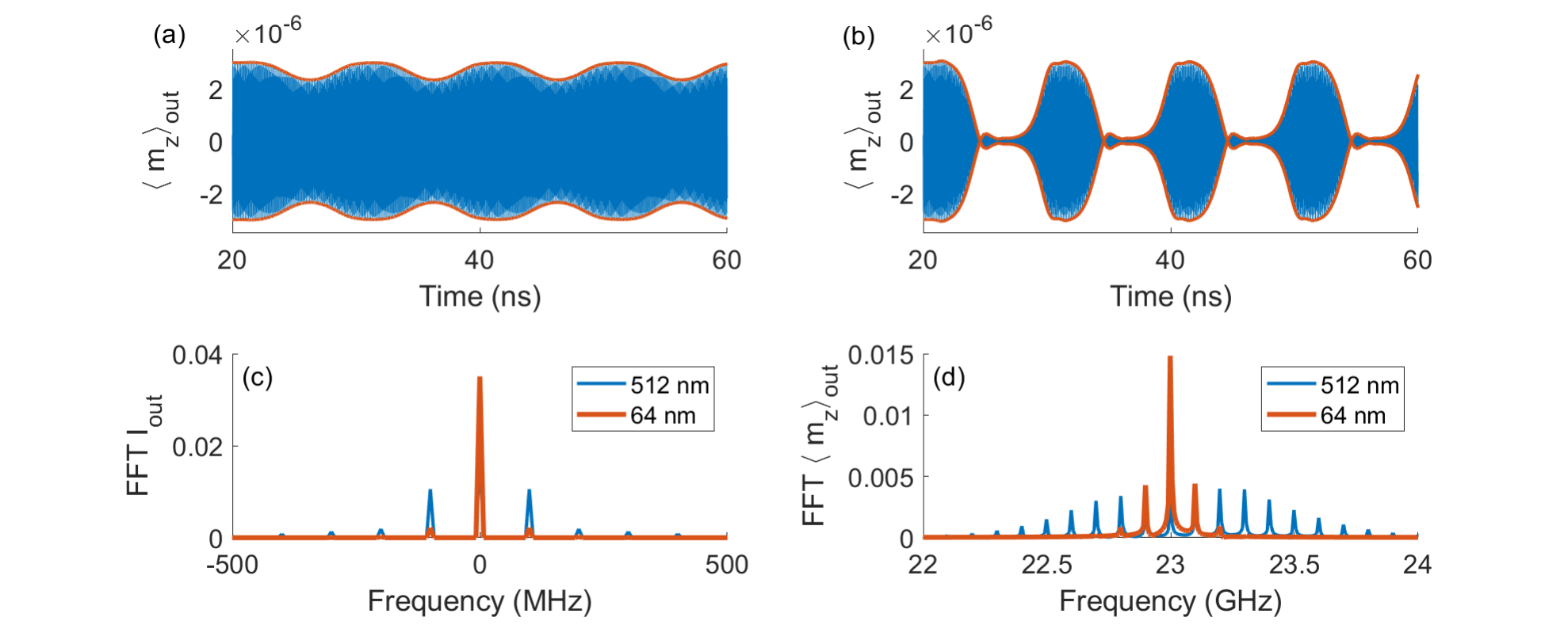}
\caption{Time-trace of the output magnetization, $\langle m_z\rangle_{out}$ for the cases (a) $\Delta w=64$~nm and (b) $\Delta w=512$~nm. In (a), the small attenuation leads to a linear modulation of the magnon amplitude. In (b), the strong attenuation leads to a nonlinear modulation of the magnon amplitude, resulting in wave-packets. In both (a) and (b), the enveloped is plotted as solid red curves. (c) Fourier transform of the envelopes of both linear modulation and wave-packets. (d) Fourier transform of $\langle m_z\rangle_{out}$ showing a predominantly amplitude modulation in the $\Delta w=64$~nm case and a frequency comb in the $\Delta w=512$~nm case. }
\label{fig:ModulatedBarrier}
\end{center}
\end{figure*} 
 
\section{Magnonic wave packet}

The field tunability of the barrier directly allows for a temporal manipulation of the band-gap. In this section, we utilize a time-varying field to produce a magnon wave packet. For simplicity, we consider a pure tone for the field, so that
\begin{equation}
    \label{eq:field}
    B_d(t) = B_{max}\left[1-\cos{(2\pi f_m t)}\right],
\end{equation}
where $B_{max}=0.04$ is the maximum field modulation and $f_m=100$~MHz is the modulating frequency. With this approach, we modify the field in the barrier from $0.05$~T to $0.15$~T. As in the previous section, we excite a magnon with a frequency of $23$~GHz and allow it to stabilize for $20$~ns.

We first set $\Delta w=64$~nm, in which case the attenuation is approximately $20$~\%. The resulting magnetization time-trace $\langle m_z\rangle_{out}$ is shown in Fig.~\ref{fig:ModulatedBarrier}(a), exhibiting a periodic modulation of the signal. The envelope is visualized with solid red curves, clearly exhibiting a sinusoidal shape. This implies that the barrier-induced magnon modulation is in a linear regime. Extending the barrier to $\Delta w=512$~nm allows us to enter the nonlinear regime. The magnetization time-trace $\langle m_z\rangle_{out}$ of this case is shown in Fig.~\ref{fig:ModulatedBarrier}(b), exhibiting magnon wave packets. In this case, essentially half of the cycle is within the band-gap so that the wave-packets are well-defined.

The linearity and nonlinearity of these cases can be better observed in Fourier space. The spectrum of the wave envelopes is shown in Fig.~\ref{fig:ModulatedBarrier}(c). For the $\Delta w=64$~nm case, the envelope only displays significant peaks at $100$~MHz, as expected for a sine wave. Higher harmonics are also observed since the simulation is not completely linear, but the second harmonic is approximately a factor 10 smaller than the fundamental peak. In contrast, the nonlinear case at $\Delta w=512$~nm exhibits a spectrum consistent with a periodic soliton pattern: a discrete Fourier transform with the envelope profile encoded in the infinite harmonic peaks. These features are reflected in the spectral characteristics of the outgoing magnon, measured with $\langle m_z\rangle_{out}$ and shown in Fig.~\ref{fig:ModulatedBarrier}(d). Both spectra are centered at the ``carrier'' frequency of $23$~GHz. The linearly modulated magnon exhibits the characteristic spectrum of amplitude modulation with two sidebands at the modulation frequency. In contrast, the nonlinear wave packet exhibits a frequency comb with as many as 19 sidebands easily observed in the linear scale.
 
\section{Conclusions}

Magnonic heterojunctions have been demonstrated numerically on the basis of field-dependent band-gap tunability. The band-gap is made possible by nano-magnonic crystals, in this case, a periodic modulation of the exchange constant in the surface wave geometry. Because of the possibility of actively changing the relative band-gap location by altering the frequency, we demonstrated that a modulated external field can, in turn, modulate magnons in both the linear and nonlinear regime. The latter achieves wave packets that exhibit a comb spectrum.

While our simulations are idealized, especially in the choice of a sharp boundary between regions of distinct magnetic field, the features shown here should be maintained in the presence of a realistic localized magnetic field. This is because the band structure requires a homogeneous region of the field to be formed. A smooth transition to this homogeneous field would only increase the size of the magnetic nano-magnonic crystal under study. A more significant problem is the magnetic damping, so that magnons can propagate through the nano-magnonic crystal and be detectable. Localized fields oriented in the plane could be produced both by a permanent magnet with an adjustable relative height to the magnetic nano-magnonic crystal or by patterning metallic antennas with fields generated by current.

We presented a simple example of magnon modulation by a dynamic band-gap where the field changes in a sinusoidal pattern. However, the field can present an arbitrary temporal function. In the nonlinear regime, for example, a train of inverse Gaussian pulses can be used to produce frequency combs with many more ``teeth'', resembling optical frequency combs. For example, such a freuquency comb could be potentially useful for a spin-wave-based spectrometer. In general, wave-packets can be used to produce a source of magnon excitations that can impinge on another material. For example, it could be fundamentally interesting to explore the collision between magnonic wave packets. Finally, the band-gap is only one aspect of the system that can be used. If operated at larger frequencies, the field will simply induce a change in the propagating frequencies to that frequency modulation could be achieved.

While the suggested possibilities are based on numerical results, nano-magnonic crystals are promising for on-device manipulation of magnons. Further experimental developments in TNL and probably the material control of low-damping magnets will aid in the development of these structures.

\section*{Acknowledgments}

This material is based upon work supported by the National Science Foundation under Grant No. 2205796.

\end{document}